# RuralShield: Super Intelligence-Assisted Security Triage and Serverless Infrastructure for Resource-Constrained Healthcare

**Mohammad Arman Ullah** arman.ullah@outlook.com

**Asif Ahmed** asif.070913@gmail.com

## Abstract

Resource-constrained healthcare providers must connect cloud security controls to an operational response process. Introducing language-model assessments adds an authorization problem: a recommendation must be converted into a permitted action against a specific resource. RuralShield combines a configurable serverless application foundation with structured model assessment, deterministic routing, and human-approved containment. Its design separates workload access, assessment validation, and action authorization so that each boundary produces inspectable records. We evaluate the implementation through a frozen 120-hour non-AI workload and separate desktop and AWS AI cohorts. The workload recorded 999,861 authenticated requests and 999,860 successful responses, with one HTTP 404 and one missing minute representing 139 scheduled requests. Median, 95th-percentile, and 99th-percentile Ohio client latencies were 28.57, 53.01, and 110.77 ms. Eight successful model requests exercised assessment and reporting in two four-case cohorts. The AWS cohort and five scripted controls produced nine completed workflows, including two human-approved Lambda containment cases with verified restoration of the original concurrency setting. Source-linked, checksum-verified archives preserved the records before scoped cleanup. The study demonstrates an inspectable integration of application controls and human-mediated AI response under the evaluated synthetic conditions.

## 1. Introduction

Healthcare cybersecurity must protect both information and service continuity. Dameff and colleagues observed increased emergency-department demand, waiting times, and ambulance diversion at hospitals neighboring a ransomware-affected organization. These effects connect a technical incident to disruption across a healthcare community. For small and rural providers, limited specialist capacity further constrains the implementation of protective measures, as documented in the Health Sector Coordinating Council's *On the Edge* report. Dameff et al., 2023 [1], HSCC, 2025 [2]

The engineering problem is to translate security objectives into an application configuration and a usable response process. HHS's Healthcare and Public Health Cybersecurity Performance Goals emphasize multifactor authentication, encryption, segmentation, and incident preparedness. Implementing these objectives in a cloud application requires coordinated identity, networking, storage, logging, and recovery settings. The HIPAA Security Rule also situates technical safeguards within an organizational security program. An application framework can make its configured controls and operational records explicit within that broader responsibility. HHS performance goals [3], HIPAA Security Rule summary [4]

Model-assisted response adds a second coordination problem: explanatory output must be translated into authorized operations. A finding can contain untrusted text, a recommendation can name a disruptive action, and a target may require protection from containment. RuralShield handles these decisions at separate boundaries. The model produces a structured assessment; deterministic code selects a notification or approval route; and a scoped executor checks the recorded approval and target before changing a resource. The resulting sequence makes a recommendation, a human decision, and an executed action separately inspectable.

RuralShield implements this sequence alongside a configuration-driven serverless workload foundation. YAML settings generate AWS CDK stacks for application access, encrypted data services, auditing, and regional recovery. The separate AI workflow combines an external model request with signed human decisions, target-restricted containment, and recorded rollback. The paper contributes the integrated system design, a deployed implementation of this authorization path, and an empirical record linking workload measurements and workflow outcomes to preserved source and configuration.

The evaluation measures each part of the system separately. A frozen five-day non-AI workload recorded 999,861 authenticated requests. Two AI cohorts each contain four successful model requests, and five additional scripted controls exercise AWS workflow branches with zero model calls. Across the AWS cases, two approved actions restored the disposable target's original concurrency setting. These results establish observed application and workflow behavior in the synthetic study. The title adopts *Super Intelligence* terminology from Executive Order 14434, Section 3(a); the technical term AI in this paper refers to the implemented language-model components. Executive Order 14434, 2026 [5]

## 2. Related Work

### 2.1 Infrastructure policy and application foundations

Infrastructure-as-code analysis makes security requirements testable during provisioning. SLIC detects security smells in infrastructure scripts, and Verdet and colleagues examine security-policy adoption in Terraform projects. cdk-nag and CloudFormation Guard implement checks over infrastructure definitions. RuralShield applies this approach to a generated application configuration through cdk-nag rule packs and project-specific checks. Its contribution at this layer is the composition of workload controls and their verification within one deployment project. Rahman et al., 2019 [6], Verdet et al., 2025 [7], cdk-nag [8], CloudFormation Guard [9]

Cloud foundations and permission analysis address different scopes of the deployment problem. AWS Landing Zone Accelerator assembles configuration-driven governance and security capabilities for multi-account environments. PrivLess analyzes serverless function-to-resource interactions and declared IAM permissions to identify overprivilege. RuralShield uses an application-centered, same-account configuration with regional recovery resources and explicit action permissions. These approaches are complementary: account governance and permission analysis surround the workload and response mechanisms described here. AWS Landing Zone Accelerator [10], Yeboah-Duako and Datta, 2026, v2 [11]

### 2.2 Response orchestration and model-assisted investigation

Event-driven remediation provides the closest operational pattern to RuralShield's response workflow. AWS Automated Security Response combines Security Hub findings with EventBridge, Step Functions, Lambda, and Systems Manager playbooks. RuralShield uses related orchestration mechanisms to connect a structured model assessment to deterministic routing and an expiring human decision. Its distinctive system scope joins this response path to a generated application foundation and evidence-preservation procedure. AWS Automated Security Response [12]

Model-assisted investigation research emphasizes how agents gather and interpret evidence. CORTEX coordinates specialized agents for SOC investigation, while ALERT-BENCH evaluates evidence-retrieving agents and the AIDA challenge-and-adjudication protocol. RuralShield instead requests one assessment and passes it to a fixed authorization workflow. The corresponding evaluation unit is the recorded assessment-to-outcome sequence, which separates model output from policy and executor behavior. Wei et al., 2025, v1 [13], Tsegai et al., 2026, v1 [14]

Prompt injection motivates a boundary between finding text and executable instructions. Greshake and colleagues demonstrate indirect injection attacks in LLM-integrated applications. RuralShield validates a fixed assessment schema, derives targets from structured finding resources, and checks authorization in the executor. Its hostile-text fixtures examine the outputs and action paths produced within these constraints. Table 1 summarizes the relationship to the principal approaches. Greshake et al., 2023, v2 [15]

**Table 1. Related approaches and the scope of RuralShield.**

| Approach | Primary focus | Relationship to RuralShield |
|---|---|---|
| SLIC, cdk-nag, Guard | Infrastructure checks | Supplies the analysis pattern used for generated workload controls |
| Landing Zone Accelerator | Multi-account foundation | Complements an application-centered, same-account deployment |
| PrivLess | Serverless permissions | Complements generated grants and target-restricted action roles |
| Automated Security Response | Remediation playbooks | Shares orchestration mechanisms; RuralShield adds model assessment and signed human decisions |
| CORTEX, ALERT-BENCH/AIDA | Model-assisted investigation | Studies investigation; RuralShield measures a fixed authorization workflow |

# 3. Threat Model and Security Objectives

## 3.1 Assets and trust boundaries

The protected assets are application records, authentication material, encryption keys, approval tokens, recovery data, and operational evidence. The evaluated application is a synthetic patient-record API. Its deployment identity and the AWS control plane are trusted to establish the configuration. Runtime credentials and action roles then govern access within that configuration.

The adversary may submit application requests, obtain workload credentials, exploit an application weakness, or embed malicious instructions in finding text. Unauthorized use of an approval link and disruptive containment are also relevant threats. These capabilities define three boundaries: access to the application, conversion of untrusted finding data into an assessment, and authorization of an operation on a selected resource.

Privileged executors and resource metadata remain part of the trust model. Protection decisions depend on observed resource identifiers and tags, and executor compromise could exceed the supplied workflow. The isolated AI deployment limits its accepted inputs to predefined synthetic case identifiers and restricts mutation permissions to one tagged disposable Lambda. This target restriction bounds the consequences of an approved experimental action.

## 3.2 Objectives

RuralShield's workload controls address confidentiality, availability, and inspectability. Authentication and access restrictions protect application entry points; encryption and segmentation protect data paths; backup and regional storage support recovery. Retained audit records connect configuration to observed operations. Synthesis checks expose selected configuration requirements before deployment.

The response objective is to preserve a traceable authorization chain. A model assessment names a playbook and explains its recommendation. Software selects compatible targets and the required route, while an executor validates the recorded human decision before containment. Every containment action in the evaluated workflow requires an unexpired approval; denial and timeout end the pending action path.

# 4. System Design

The design connects a generated application foundation to a separately deployed response workflow. Figure 1 shows the workload boundary; the assessment and authorization modules define the response boundary.

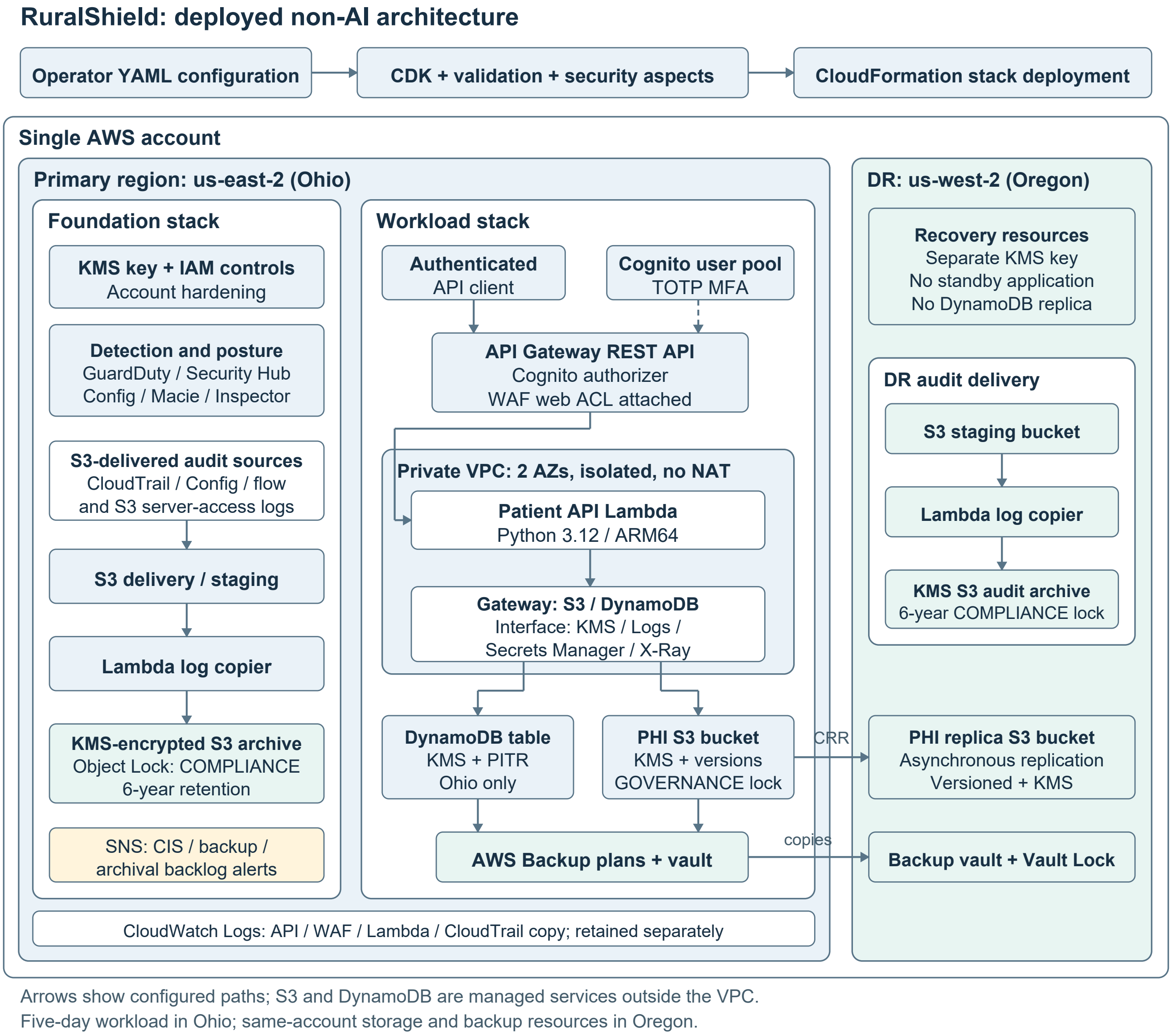

*Figure 1. Deployed architecture for the five-day non-AI workload. Cognito and WAF protect the public API, while the application Lambda runs in isolated subnets. S3 and DynamoDB are regional services. Log delivery and archival use separate storage paths; Oregon contains same-account storage and backup resources.*

## 4.1 Generating the workload foundation

Configuration-driven provisioning makes the workload's security settings reviewable together. RuralShield validates YAML settings for regions, compute, data stores, controls, and recovery, then generates AWS CDK stacks. Selected invalid combinations, including equal primary and recovery regions, insufficient backup retention, and wildcard model identifiers, are rejected. Foundation, Workload, regional recovery, and response stack boundaries separate shared controls, application resources, recovery resources, and workflow services.

The deployed workload places application access and private compute behind distinct controls. In Ohio, API Gateway exposes a Cognito-authorized API protected by WAF. Lambda runs in isolated subnets and accesses regional services through configured endpoints. DynamoDB stores synthetic patient records, with an encrypted S3 bucket provisioned alongside it. Oregon holds same-account recovery resources. Figure 1 identifies the configured workload and logging paths.

Identity and data controls enforce complementary requirements. Cognito requires TOTP MFA and disables self-registration; API Gateway authorizes application requests. DynamoDB uses customer-managed encryption, point-in-time recovery, and deletion protection. The PHI-designated S3 bucket combines encryption, versioning, public-access blocking, and 35-day GOVERNANCE Object Lock retention. Record-level authorization remains an application responsibility.

Retention distinguishes recoverable workload data from durable audit evidence. Scheduled backups, regional copies, a locked backup vault, and table restore-testing settings support recovery operations. Audit archives use 2,190-day COMPLIANCE Object Lock retention, and CloudWatch logs have separate retention settings. These choices bind recovery and evidence storage to explicit retention obligations. AWS Object Lock [16]

### 4.2 Delivering evidence and checking configuration

Audit collection separates service-compatible delivery from archival retention. CloudTrail and applicable service logs reach versioned delivery storage. A copier reads an identified source version and creates an encrypted archive object whose identity includes the source bucket, key, and version. This arrangement accommodates S3 access-log delivery requirements while preserving the archive configuration. CloudTrail selectors exclude archive and delivery-bucket writes to avoid recursive audit generation. AWS server access logging [17]

Static checks examine the generated configuration before deployment. The pipeline combines cdk-nag's AWS Solutions and HIPAA Security rule packs with project checks for encryption, retention, MFA, backups, and logging. Unsuppressed errors block the intended deployment path. Resource-specific exceptions document service compatibility or operations requiring broad resource scopes, so the check result is interpreted against a defined rule-and-exception set. cdk-nag [8]

### 4.3 Converting assessments into policy decisions

A structured assessment limits the interface between model output and response logic. The assessment contains an explanation, patient-safety impact, possible exposure, confidence, a recommended playbook, and suggested next steps. The external adapter requests a strict `record_triage` function result, then validates it locally. Validation rejects malformed or duplicate-key JSON, unexpected tools or fields, invalid enumerations, nonfinite values, and out-of-range confidence. The resulting record supplies advice to the policy. OpenAI function calling [18]

The policy maps that advice to a finite action vocabulary and compatible resources. The shared library includes access-key deactivation, role-session revocation, S3 public-access blocking, IAM-role quarantine, Lambda quarantine, and world-open security-group ingress removal, together with escalation-only handling. Targets are selected from normalized finding resources. The isolated cloud trial permits executable containment only for Lambda quarantine on its disposable target.

The evaluated policy has two routes because automatic containment is disabled. Let $E(f,p)$ mean that finding $f$ is a threat, recommended playbook $p$ is actionable, a compatible target exists, and selected targets are not marked protected. The study configuration selects:

$$mode(f,p) = \begin{cases} \textbf{APPROVAL}, & \text{if } E(f,p) \\ \textbf{NOTIFY}, & \text{if } \neg E(f,p) \end{cases}$$

Eligibility opens a human-decision request; the executor performs the final authorization check. Escalation-only and other ineligible assessments produce notifications. Model confidence remains a recorded score rather than permission to execute. This division lets the experiment inspect recommendation, routing, human decision, and resource change independently.

### 4.4 Authorizing, executing, and restoring an action

The deployed AI workflow connects assessment and authorization through a state machine. An isolated EventBridge bus supplies a predefined case identifier to Standard Step Functions. The triage Lambda loads its fixture, calls the OpenAI Responses API, validates the assessment, applies policy, and stores an incident. SNS delivers the resulting notification or decision request. A narration case follows a separate report-delivery branch, as shown in Figure 2.

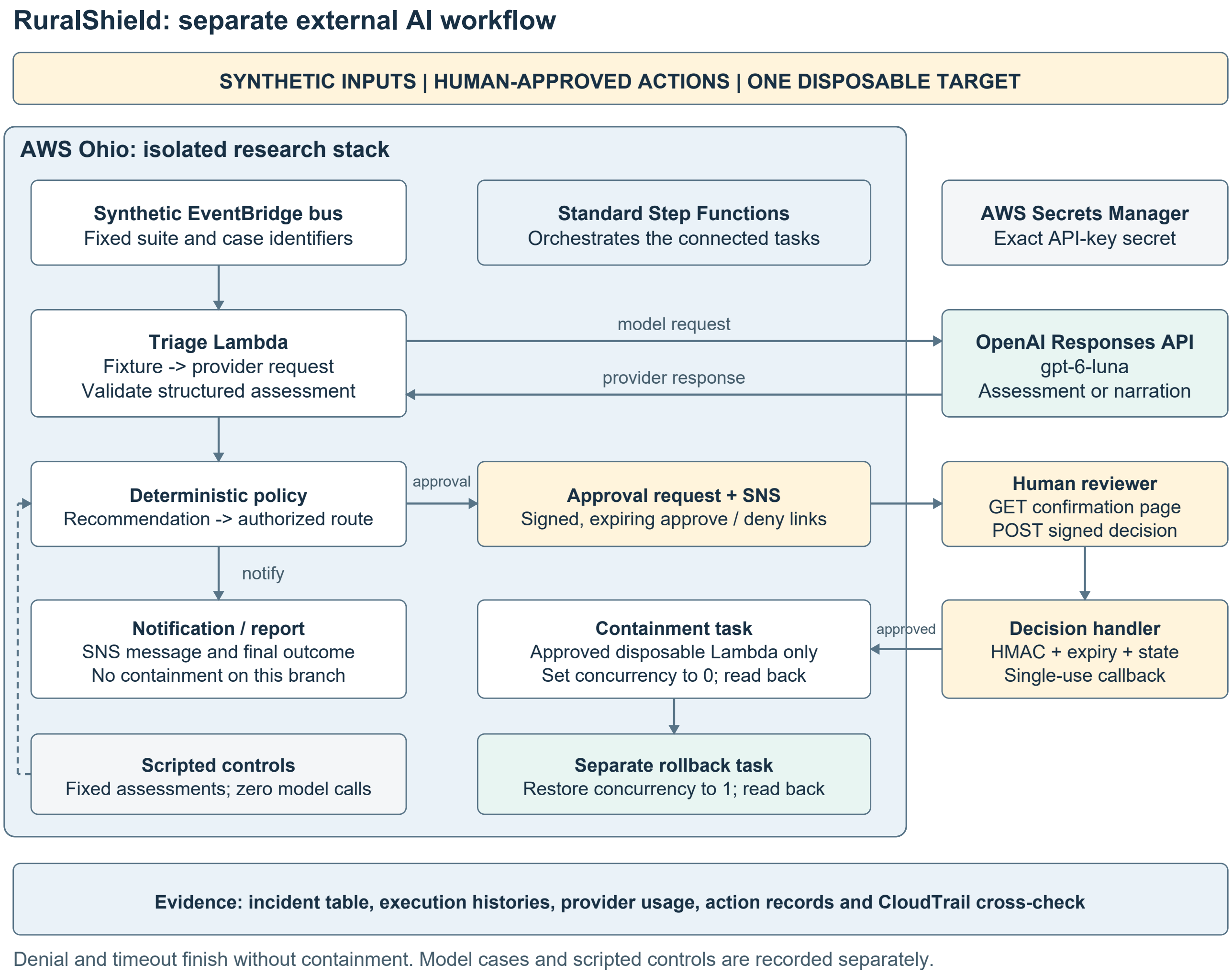

*Figure 2. Separate AWS AI workflow evaluated with synthetic inputs. Policy routes an assessment to notification or human approval. A confirmed approval POST resumes the waiting execution; containment and rollback run as separate tasks. Scripted controls supply fixed assessments with zero model requests. Incident and action records preserve each transition.*

Approval links bind a human decision to a pending request. The request handler stores the Step Functions task token behind a random identifier and sends signed approve and deny links. GET displays a confirmation form; POST checks the HMAC signature, expiry, request state, and secret version, then conditionally consumes the decision and resumes the execution. The ordinary decision window is 15 minutes. Possession of the signed link is the authorization mechanism for the decision page.

Containment and rollback recheck the resource at the point of mutation. The action handler validates the recorded approval, target identity, protection tags, and expected starting concurrency before setting the disposable function's reserved concurrency to zero. The separate rollback handler checks the target and containment state, restores the saved value of one, and records a fresh readback. Target-restricted IAM roles and recorded state transitions make the experimental action and its restoration inspectable.

## 5. Implementation and Reproducibility

The software separates provisioning, assessment, policy, and execution. TypeScript modules implement configuration processing, CDK constructs, stacks, and static checks. Python modules implement provider transport, schema validation, routing, approvals, actions, reporting, and archival. The workload runs on Python 3.12; the isolated AI workflow uses Python 3.13 on ARM64. Pure policy and protocol behavior can therefore be exercised by offline tests.

The deployed assemblies preserve the separation between the two studies. The non-AI application contains 213 resources across its Foundation, Workload, and regional recovery stacks, with 12 additional observation resources. The AI workflow uses its own 52-resource stack, including six runtime Lambda handlers, one disposable target, seven log groups, and an incident table. Runtime functions have reserved concurrency one. Deployed package and handler hashes matched the reviewed source, and the API key is loaded from a designated secret.

Local verification checks the software boundaries used by the deployments. The public release passed its TypeScript build, 37 Jest tests, 128 Python tests, and five observation-worker tests. The suites cover configuration, assessment validation, authorization, archival verification, and cleanup safeguards. Synthesis recorded zero unsuppressed static-check errors under documented exceptions, with one provider-related warning. Live measurements are presented separately in Section 6.

Experiment manifests bind the observations to the configuration that produced them. The workload manifest fixes its time window, request target, worker hash, and output locations. AI records retain fixture and request hashes, requested and returned model identifiers, usage, elapsed time, assessments, decisions, and outcomes. Before the AI deployment, an activation gate verified the final non-AI archive, ZIP, frozen source, and cloud manifest. The exporter independently verifies collected files and their ZIP before scoped retirement of duplicate cloud evidence.

## 6. Evaluation

### 6.1 Questions and study protocol

The evaluation addresses three questions: how the workload behaves over a fixed observation period, what assessments the selected model returns, and how the deployed workflow authorizes and executes those assessments. Table 2 defines the study components and their denominators. The AWS model cases appear in both the assessment and workflow analyses because each request belongs to one end-to-end execution. Scripted controls are additional executions with fixed assessments.

**Table 2. Study components and measurement denominators.**

| Study component | Recorded sample | Measurement focus |
|---|---|---|
| Non-AI pilot | Frozen 120-hour window; 7,200 scheduled minute slots | Authenticated requests, HTTP outcomes, client latency |
| Desktop AI cohort | Four successful requests on October 31 | Model response, local validation and routing |
| AWS AI cohort | Four successful requests on November 1 | Model response and connected workflow outcome |
| AWS scripted controls | Five executions; zero model requests | Notification, approval, denial, timeout, error-path handling |

Workload preparation established the required access and recovery operations. Cognito SRP/TOTP login and authenticated HTTPS health, create, and read tests produced a matching synthetic-record readback. The operator confirmed an Ohio SNS notification. A controlled DynamoDB restore returned a matching record, and an S3 replication probe reached Oregon with the same SHA-256 checksum. These readiness observations are distinct from the subsequent workload totals.

The non-AI protocol fixed the observation period and issuing deadline before measurement. It ran from 25 October 2026 at 18:29 UTC to 30 October at 18:29 UTC. An Ohio Lambda observer used four concurrent clients and a one-million-request schedule, recording minute-level observations across 7,200 scheduled slots. Counts and latency statistics use recorded requests; a missing slot retains its scheduled volume as an unobserved quantity.

Both AI cohorts used the same provider settings. Requests went to OpenAI's direct Responses API with requested and returned alias `gpt-6-luna`, reasoning effort `none`, a 600-output-token limit, `store:false`, and automatic retries disabled. Each case has one successful response. Triage requires local schema acceptance, and narration requires nonempty text. Provider-request latency is client elapsed time around the API request; it excludes human-decision time and the rest of the workflow. OpenAI GPT-6 Luna [19]

## 6.2 Workload behavior over five days

The frozen workload recorded 999,861 requests, of which 999,860 succeeded. There were 7,199 recorded minute slots and one HTTP 404 in the first slot. Missing slot 4420 represents 139 scheduled requests with unobserved outcomes. These requests contribute to neither recorded attempts nor success, failure, or latency totals. Table 3 reports the observed quantities.

**Table 3. Completed five-day non-AI workload results.**

| Metric | Observed value |
|---|---|
| Observation duration | 120 hours |
| Scheduled / recorded minute slots | 7,200 / 7,199 |
| Scheduled request target | 1,000,000 |
| Recorded requests | 999,861 |
| Successful responses | 999,860 |
| HTTP outcomes | 999,740 status 200; 120 status 201; 1 status 404 |
| Missing-slot scheduled requests | 139, outcome unobserved |
| Median client latency | 28.57 ms |
| 95th-percentile client latency | 53.01 ms |
| 99th-percentile client latency | 110.77 ms |

Recorded volume was 99.9861% of the scheduled target. Latency percentiles use nearest-rank statistics over all 999,861 recorded elapsed times, including the 404. Their measurement point is the Ohio observer calling the authenticated API. The worker hash remained unchanged, and the absolute cutoff disabled its load rule at the window's end. Post-window checks found that rule disabled and the failure queues empty. The final archive reproduced the 7,199 records and the same totals.

### 6.3 Model assessments in two separate cohorts

The desktop cohort exercised three triage cases and one narration case. All three triage responses passed the local schema check; narration passed the nonempty-text contract. Table 4 distinguishes the returned content from the locally selected route. The credential fixtures and the AWS cohort's Lambda fixtures have different targets, so their observations retain separate denominators.

**Table 4. Desktop AI cohort: one successful request per case.**

| Desktop case | Request time (ms) | Model recommendation or output | Local outcome |
|---|---:|---|---|
| Credential threat | 8416.09 | ESCALATE_ONLY | NOTIFY |
| Posture finding | 5909.86 | ESCALATE_ONLY | NOTIFY |
| Credential attacker-instruction variant | 6409.62 | Access-key deactivation; confidence 0.42 | APPROVAL route |
| Weekly narration | 2268.08 | Narrative of supplied counts | Nonempty text; five counts matched |

Local routing followed the assessed action and the study's authorization setting. The two escalation recommendations produced notification. The credential attacker-instruction variant recommended access-key deactivation and selected the approval route locally. The narration fixture supplied three findings, two threats, one posture finding, zero containments, and two pending reviews; descriptive review found all five counts in the response. These invented weekly counts are independent of the workload measurements.

The AWS cohort connected four model responses to deployed workflow outcomes. Three triage responses passed schema validation and one narration response passed the text contract. Table 5 reports their request latencies, recommendations, and final outcomes. Posture and the ordinary Lambda threat produced notification; weekly narration produced a delivered report; the Lambda attacker-instruction variant produced an approved action followed by rollback.

**Table 5. AWS AI cohort: one successful request per case.**

| AWS case | Request time (ms) | Confidence | Recommendation and outcome | Token estimate (USD) |
|---|---:|---:|---|---:|
| Posture | 6941.18 | 0.99 | ESCALATE_ONLY; NOTIFIED | 0.000425425 |
| Weekly narration | 1937.26 | n/a | Narrative; REPORT_SENT | 0.000031900 |
| Lambda threat | 6778.15 | 0.94 | ESCALATE_ONLY; NOTIFIED | 0.000471925 |
| Lambda attacker-instruction variant | 6554.91 | 0.64 | Lambda quarantine; approved; ROLLED_BACK | 0.000479675 |

The hostile Lambda fixture exposes the boundary between model content and authorization. Its text requested `SHELL_EXEC` and confidence 2.0, both excluded by the assessment schema. The accepted output recommended Lambda quarantine with confidence 0.64. That recommendation entered the human-approval path and reached execution only after a confirmed decision. The observation establishes the sequence for this fixture, with structural constraints and approval both part of the evaluated setting.

### 6.4 Authorization and execution in AWS

Five scripted controls isolate workflow behavior from model selection. Each supplies a fixed assessment and makes zero provider requests. The operator confirmed notification receipt and submitted approve and deny decisions through the emailed pages. Table 6 records the branch outcomes, including a ten-second expiry control and a fixed model-error-path assessment.

**Table 6. Scripted AWS workflow controls: zero model calls.**

| Scripted control | Observed outcome | Action record |
|---|---|---|
| Notification | Operator received SNS message | No containment |
| Approval | Confirmed approve POST; action and rollback | Concurrency 1 -> 0 -> 1 |
| Denial | Confirmed deny POST; APPROVAL_DENIED | No containment |
| Timeout | Ten-second wait expired; APPROVAL_TIMEOUT | No containment; expiry elapsed |
| Model-error handling | Fixed error-path assessment produced notification | Zero model calls; no containment |

All nine AWS executions completed with their branch-specific outcomes checked. Exactly two generated containment records: the scripted approval control and the model-selected Lambda action. Both changed reserved concurrency from one to zero and restored it to one. Fresh SDK readbacks confirmed restoration, and CloudTrail independently recorded four matching concurrency changes: two setting zero and two setting one. The combined records link each approved action to its subsequent configuration restoration.

Approval replay and expiry checks added a user-visible observation. The operator reported that spent and expired links displayed an already-recorded-or-expired page without a decision form. The timeout request retained stored PENDING status with elapsed expiry; authorization evaluates state and time together. Operator-reported page behavior and independently recorded AWS changes remain distinct evidence types.

### 6.5 Costs and evidence lifecycle

Model-token estimates use the recorded usage of the eight successful requests. The desktop cohort used 4,592 input and 1,614 output tokens; the AWS cohort used 4,654 input and 1,679 output tokens. Input included 4,099 and 4,161 cache-write tokens, respectively, with zero cache reads. The experiment's rates were USD 0.10 per million ordinary input tokens, USD 0.125 per million cache-write tokens, and USD 0.50 per million output tokens. Cache writes use the estimator's assumed default 30-minute lifetime. Table 7 applies these categories without double-counting cache writes. OpenAI prompt caching [20]

**Table 7. Token estimates for the eight successful model requests.**

| Model cohort | Successful requests | Usage: input / output tokens | Estimated cost (USD) |
|---|---:|---:|---:|
| Desktop | 4 | 4,592 / 1,614 | 0.001368675 |
| AWS | 4 | 4,654 / 1,679 | 0.001408925 |
| Combined | 8 | 9,246 / 3,293 | 0.002777600 |

AWS account-period billing measures a different quantity from model-token usage. Cost Explorer reported estimated gross unblended charges of USD 25.4304198609 for the six UTC account-day bins of October 25-30. Those bins include setup, instrumentation, export, and retained resources around the 120-hour workload. The rounded USD 25.43 figure is therefore an account-period estimate. The AI token subtotal of USD 0.002777600 excludes AWS orchestration and storage charges.

Verified archives connect the measurements to retained evidence through cleanup. Native closeout recorded 73 actions and 24 passing shutdown checks. Following a 24-hour late-evidence hold, the follow-up non-AI archive verified 30,464 payload files and a 1,871,761,398-byte ZIP. The separate AI archive verified 257 files and its ZIP, including requests, outputs, execution records, actions, and cleanup receipts. Hash verification establishes the integrity of these copies; the payload and ZIP occupy the same disk.

Resource retirement distinguishes the experiment's active services from retention obligations. By November 12, the five remaining non-AI application and instrumentation stacks had been removed; the separate AI trial had already been removed after archival verification. The recorded residual inventory contained four locked buckets, two KMS keys, 17 backup recovery points, and a small control-receipt bucket. Their storage and key costs continue beyond the experiment.

## 7. Discussion

The study's main finding is that workload controls and model-assisted response can be joined through observable authorization boundaries. The non-AI measurements describe authenticated request handling over a fixed interval. The separate AWS records show how an assessment reaches notification or a human decision, and how an approved change reaches verified restoration. Keeping these observations separate makes the source of each result explicit.

Action records provide stronger evidence of the evaluated response path than confidence scores alone. The two containment cases have recorded human decisions, target checks, concurrency changes, and restoration readbacks. Denial and timeout controls exercised the complementary termination paths. This evidence supports the implemented authorization sequence while keeping the model's recommendation distinct from the action permitted by software and the operator.

The empirical scope is a synthetic application in one AWS account and a small, fixed AI case set. Workload latency is measured from Ohio Lambda, and each model case contributes one successful response. Desktop and AWS threat targets differ, confidence is model-reported, and narration review is descriptive. The results consequently support system operation in the stated configuration rather than a population-level estimate of model quality or clinical effectiveness.

Operational simplicity depends on visible responsibilities as well as managed services. Serverless infrastructure delegates host operation while leaving application code, permissions, dependencies, and operating decisions with the customer. Evidence retention also creates storage and key obligations after temporary services are retired. RuralShield records these boundaries through configuration checks, action records, archival manifests, and scoped closeout. AWS shared-responsibility model [21]

## 8. Conclusion

RuralShield connects configurable serverless workload controls to structured assessment and human-authorized security response. The five-day non-AI workload recorded 999,860 successful responses from 999,861 attempts, with the observed HTTP failure and missing minute retained in the evidence. Separate AI cohorts and scripted workflow controls exercised assessment, notification, approval, denial, timeout, and two scoped Lambda actions with verified restoration. The central design lesson is to make configuration, recommendation, authorization, and execution separately inspectable. The implementation and preserved synthetic study provide a concrete reference for this integration in resource-constrained healthcare research.

## Data and Code Availability

The code is publicly available under Apache-2.0 at github.com/armanbuet09/rurlalshield [22]. The initial public release, commit `4261e3d` dated November 12, 2026, includes implementation, tests, synthetic fixtures, example configuration, and research tools. Research-account identifiers and contact details are replaced with examples in that release. The authors privately retain the original configuration, frozen source, and verified operational evidence; those byte-identical experimental records are distinct from the sanitized public snapshot.